Mixed-Effects Modeling of NYC Subway Ridership Using MTA and Weather Data

Zoe Curtis, Jake Haines

North Carolina State University

April 28, 2025

## 1. INTRODUCTION

This analysis aims to explore temporal patterns and the role of weather in average monthly ridership of New York City subways. Our dataset comes from Metropolitan Transportation Authority (MTA) [2]. We are using the data from 2023 on subway ridership and concatenated it with 2023 NYC weather data from Weather Underground [3]. The MTA dataset was built via "Destination Inference"; since NYC only tracks swipe-ins at turnstiles, they estimate the end destination with the next swipe-in from that Metrocard. By aggregating the origin-destination pairs for each hour, and then further for each month, the dataset contains monthly estimates for each possible trip at each hour of the day. Since there are 472 subway stations in NYC and thus a maximum of 111,156 possible origin-destination pairs one could take, we created a data pipeline using Apache Airflow to aggregate and transform data from multiple sources to build our sample [4].

We focus on the effects of origin borough, seasonal trends, and weather conditions over the course of 2023. Due to the large size of the MTA origin-destination dataset, we developed an ETL pipeline using PostgreSQL and Apache Airflow to efficiently prepare and manage the data, supplementing it with aggregated weather data from Weather Underground. The analysis uses longitudinal monthly data and a combination of mixed-effects models and principal component analysis (PCA) to explore relationships between ridership and external factors.

The report addresses four main questions: (1) whether a linear trend in ridership exists across the year and varies by borough, (2) whether origin borough significantly influences baseline ridership, (3) how weather variables such as gust speed affect ridership, and (4) whether seasonal effects, particularly in December, are confounded by weather conditions. Overall, the analysis finds that origin borough, especially Manhattan, is associated with differences in average ridership, and that maximum gust speed is a significant factor influencing monthly variation.

The remainder of the report is structured as follows: Section 2 contains data preparation, exploratory analysis, and model development; Section 3 presents the main results and interpretations; and the conclusion discusses key findings, limitations, and directions for future work.

## 2. METHODS

### 2.1 Data Preparation
To begin the analysis, we had to create an ETL (extract-transform-load) pipeline to prepare our data. The 2023 MTA origin-destination dataset contained over 115 million rows and 16 columns, which made it too large to download manually, especially with MTA's user interface setting a limit of 1000 rows per export. To work around this, we accessed MTA's application programming interface (API) which allowed for more flexible and large-scale querying from their database. We also had supplementary and covariate data that was gathered from Weather Underground that needed to be aggregated by station ID and month. To handle the volume and complexity of these datasets, we housed them in a PostgreSQL database, where all transformations and joins were performed at the database level for efficiency. Apache Airflow was used to build a modular pipeline that could orchestrate these tasks in parallel and automate the sequence of API pulls, CSV loads, and database operations. A detailed overview of the pipeline structure and the structure of the final dataset is provided in the appendix.

### 2.2 Overall Ridership
Utilizing R, we created visual summaries of the overall data, included below.

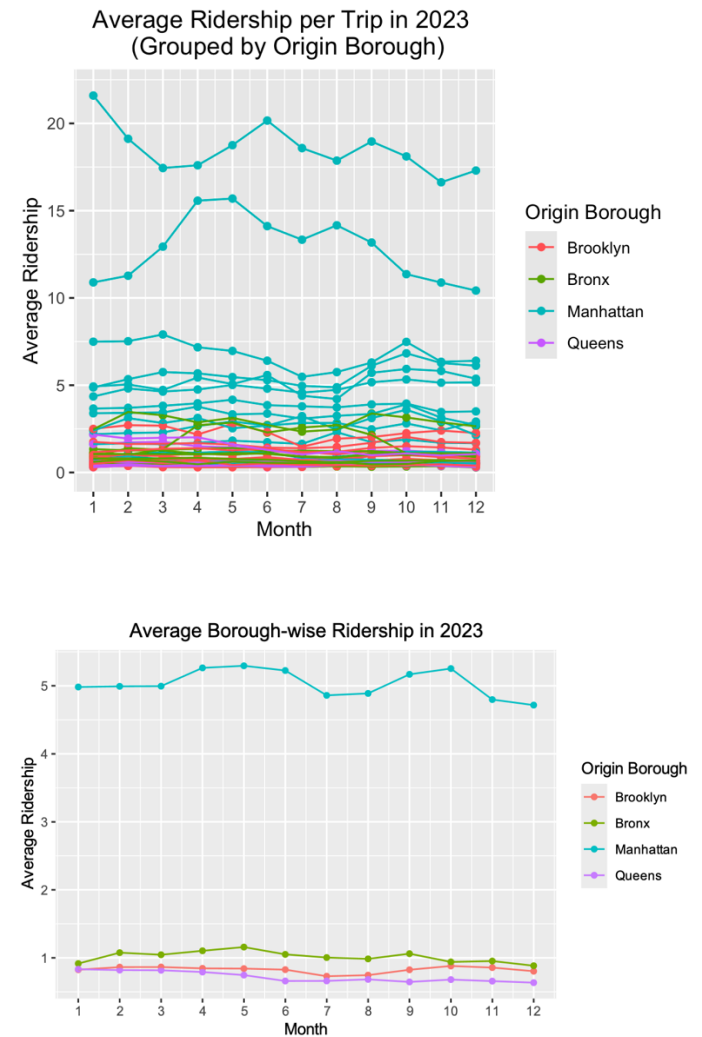

We grouped the data by origin borough, which indicated that Manhattan trips had much higher intercepts on average than the other three boroughs. We also tried grouping by destination borough, but this did not show as much of a trend. At first, visual inspection does not indicate a strong linear trend, or even a crest in summer months as one might expect.

The two trips with the highest ridership stand out quite a bit–their intercepts are higher and their trajectory shows a lot more variation than the rest of the data. Further inspection reveals that the top trip goes from Grand Central-42 St to 61 St-Woodside in

Queens. One possible explanation for the popularity of this trip is that the 61 St-Woodside stop connects with the Long Island Railroad, so it would be popular for commuters. This trip also peaks in January, which could be because during snowy weather, commuters switch from cars to public transit. The second highest trip goes from 125th street in Manhattan to 161 St-Yankee Stadium. This is a quick direct trip, and connects to Yankee Stadium, a popular destination for tourists. The baseball stadium could also explain this trip's popularity in the summer.

Considering that trips with an origin borough of Manhattan have notably different intercepts than the other boroughs, we build a model including an indicator for cases where the origin borough is Manhattan. The fixed effects include slope and intercept for Manhattan and non-Manhattan originating trips. Additionally, there appears to be higher variation in the intercepts of the trips from the Manhattan group. For this reason, we include group indicators for the random effects of intercepts. Slope does not appear to have higher variation in any group, so the random effect for slope does not have group indicators to avoid overfitting. This model is defined in the following section.

### 2.2.1 Mixed-Effects Model with Borough-Specific Fixed and Random Effects

$$\text{For } M_i = 1: \quad Y_{ij} = (\beta_{0M} + b_{0i,M}) + (\beta_{1M} + b_{1i})t_{ij} + e_{ij}$$

$$\text{For } M_i = 0: \quad Y_{ij} = (\beta_{0N} + b_{0i,N}) + (\beta_{1N} + b_{1i})t_{ij} + e_{ij}$$

where:
- $Y_{ij}$: average ridership for pair $i$ at month $j$
- $M_i$: indicator (1 if origin is "M", 0 otherwise)
- $t_{ij}$: month
- $\beta_{0M}$, $\beta_{0N}$: fixed intercepts for borough $M$ or other $N$
- $\beta_{1M}$, $\beta_{1N}$: fixed month slopes for borough $M$ or other $N$
- $b_{0i,M}$, $b_{0i,N}$: random intercept deviations for pair $i$ in borough $M$ and $N$
- $b_{1i}$: random slope deviation for month for pair $i$
- $e_{ij}$: residual error for pair $i$ at month $j$

We use different parametrization for our model, compound symmetry for the within subject correlation structure, and a D matrix with general structure for the random effects. We also tested this same model with a diagonal structure for the D matrix, but the AIC was lower for the general structure.

A few alternative models were tested in addition to the previous. One had indicators for each borough, but the data do not give us enough indication that Brooklyn, the Bronx, and Queens have different intercepts, and this created a very complicated model. However, by running the following model, we were able to test if those boroughs did have significantly different intercepts:

### 2.2.2 Preliminary Full Mixed-Effects Model for Borough-Specific Intercepts and Time Slopes

$$Y_{ij} = M_i\beta_{0M} + B_{ki}\beta_{0Bk} + B_{xi}\beta_{0Bx} + Q_i\beta_{0Q} + M_it_{ij}\beta_{1M}$$
$$+ B_{ki}t_{ij}\beta_{1Bk} + B_{xi}t_{ij}\beta_{1Bx} + Q_it_{ij}\beta_{1Q} + b_{0i} + t_{ij}b_{1i} + e_{ij}$$

The intercepts for Brooklyn, the Bronx, and Queens (denoted by "Bk", "Bx", and "Q", respectively) were not significant, so it is reasonable to proceed with the simplified model that compares Manhattan to the rest of NYC.

We also repeated the same model above but with random effects for the intercept of each borough, however, this model failed to converge in R and runs the risk of overfitting. This model is defined as:

### 2.2.3 Origin Borough Mixed Model with Borough-Specific Intercepts and Slopes

$$\begin{aligned}Y_{ij} =\ & M_i\beta_{0M} + M_ib_{0i,M} + B_{ki}\beta_{0Bk} + B_{ki}b_{0i,Bk} + B_{xi}\beta_{0Bx} + B_{xi}b_{0i,Bx}\\ & + Q_i\beta_{0Q} + Q_ib_{0i,Q} + M_it_{ij}\beta_{1M} + B_{ki}t_{ij}\beta_{1Bk} + B_{xi}t_{ij}\beta_{1Bx}\\ & + Q_it_{ij}\beta_{1Q} + t_{ij}b_{1i} + e_{ij}\end{aligned}$$

where M=1 if origin borough is Manhattan and 0 otherwise, Bk=1 if origin borough is Brooklyn and 0 otherwise, Bx=1 if origin borough is the Bronx and 0 otherwise, and Q=1 if origin borough is Queens and 0 otherwise. This model allows us to draw inferences about the role of origin borough.

## 2.3 Borough Stratification

Through multiple models, we found Manhattan-originated trips to be biasing our analysis due to the significance of borough on ridership and high overall ridership of Manhattan-originated trips. One method used to handle this was to stratify analyses on origin borough. So we subset the data into four datasets containing records from each respective borough, and fit models to them to investigate.

## 2.4 Weather Covariates

In addition to using origin borough, we wanted to build on the previous model and explore what role weather plays. The weather variables appear to change across the year, although not in a linear fashion. Our dataset contains the variables *max_temp*, *avg_temp*, *min_temp*, *max_dew_point*, *avg_dew_point*, *min_dew_point*, *total_precip*, *max_wind*, *avg_wind*, *max_gust*, and *avg_gust*.

### 2.4.1 Visual Summaries

Some variables of interest included average total precipitation, maximum gust speed, average gust speed, maximum dew point, and average dew point. We examine their trends visually:

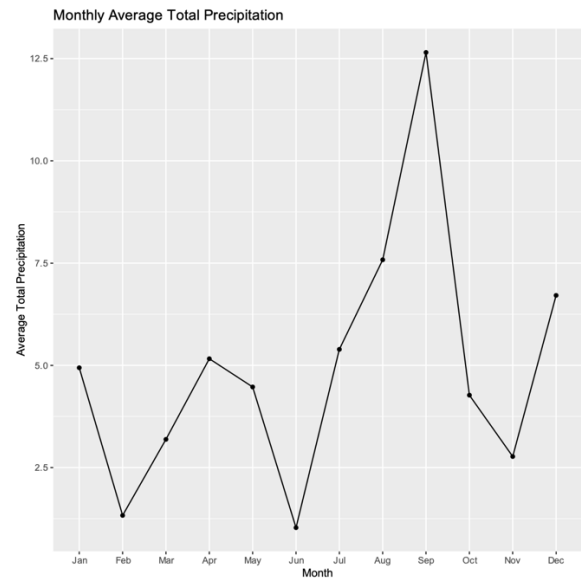

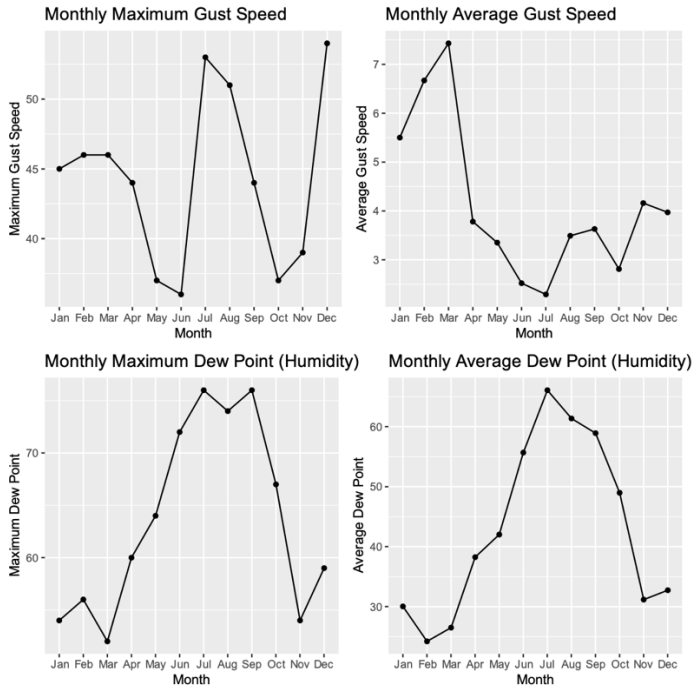

We can see the total precipitation aligns with expected seasonal weather patterns: December/January (snow), April-May (spring showers), August-October (hurricane season). Similarly, we see maximum and average dew points appear to be correlated. In addition, maximum gust speed and average gust speed do not follow the same trends on visual inspection. This indicates confounding in the behavior and seasonal patterns of gusts.

Overall, many of the weather variables are naturally collinear, so we can reduce them and identify which individual variables would be of interest for modeling, both of which can be approached using principal component analysis.

*2.4.2   Principal Component Analysis*

PCA was used to explore what covariates may be of interest for building models further. The PCA conducted produced the following loadings, which explain approximately 90% of variance in weather conditions:

| Variable | PC1 | PC2 | PC3 |
|---|---|---|---|
| max_temp | 0.34 | 0.02 | 0.28 |
| avg_temp | 0.36 | -0.03 | 0.09 |
| min_temp | 0.35 | -0.08 | 0.09 |
| max_dew_point | 0.36 | -0.10 | 0.07 |
| avg_dew_point | 0.37 | -0.08 | 0.03 |
| min_dew_point | 0.36 | -0.03 | 0.06 |
| total_precip | 0.16 | -0.49 | 0.35 |
| max_wind | -0.15 | -0.59 | 0.02 |
| avg_wind | -0.30 | -0.14 | 0.60 |
| max_gust | -0.01 | -0.58 | -0.57 |
| avg_gust | -0.33 | -0.15 | 0.31 |

We analyzed the loadings of the three most important principal components. The strong negative loadings for PC2 of total precipitation, maximum wind speed, and maximum gust speed indicate PC2 is related to stormy conditions. In PC3, the strong positive loading for average wind speed and strong negative loading for maximum gust speed indicate PC3 represents a measure of wind behavior, that is, whether wind is steady or gusty. PC1 is more broad and hence not as interpretable, as temperature and dew point (humidity) have moderate positive loadings while average wind and average gust speed have negative positive loadings of the same magnitude.

Given the loadings for the PC2 and PC3, we now have insight on what weather covariates dominate among themselves. Thus, we will explore effects of precipitation, wind speeds, and gust speeds on monthly ridership.

Additionally, we will run the previous model, but adding fixed effects for intercept and slope for the first three principal components. It should be noted that PCA was only applied to external covariates by month, which is completely exclusive of ridership (the response variable). This maintained the longitudinal structure of the data. When building models using principal components, the principal components act as time-varying predictors.

### 2.5   Effects of Weather on Ridership

*2.5.1   Baseline: Linear Mixed-Effects Model Assessing Impact of Precipitation and Wind*

To introduce weather covariates to the analysis, we started building simple models as a baseline using the weather covariates as fixed effects to examine whether any particular weather covariate has significance in modeling ridership. The baseline model built was:

$$Y_{ij} = \beta_0 + \beta_1 \text{total\_precip}_{ij} + \beta_2 \text{max\_wind}_{ij} + \beta_3 \text{max\_gust}_{ij} + u_i + \epsilon_{ij}$$

where:
- $Y_{ij}$: average ridership for pair $i$ at month $j$
- $\beta_0$: fixed intercept
- $\beta_1$: fixed intercept coefficient for total precipitation
- $\beta_2$: fixed intercept coefficient for maximum wind speed
- $\beta_3$: fixed intercept coefficient for maximum gust speed
- $u_i$: random intercept for pair $i$
- $\epsilon_{ij}$: residual error for pair $i$ at month $j$

Adding origin and destination boroughs to this model confirms the dominance of Manhattan-originated trips that we saw earlier, and interestingly retains significance of maximum gust speed. Next, we introduce origin borough $k$ as a random effect $v\_k$ to address potential bias caused by the Manhattan indicator and overall dominance in ridership.

*2.5.2   Mixed-Effects Model with Borough Random Intercepts*

$$Y_{ij} = \beta_0 + \beta_1 \text{max\_gust}_{ij} + \beta_2 \text{month}_{ij} + u_i + v_k + \epsilon_{ij}$$

*2.5.3   Stratified Borough-wise Mixed-Effects Model for Gust Speed Analysis*

To address bias present from origin borough, we stratified the data. After stratification, we tried fitting the general random effects model to each borough to examine where maximum gust speeds may impact ridership:

$$Y_{ij} = \beta_0 + \beta_1 \times \text{max\_gust}_{ij} + u_i + \epsilon_{ij}$$

### 2.5.4 Manhattan-Gust Interaction Model

To examine whether there is interaction between Manhattan-originated trips and maximum gust speed, we build a mixed model with a dummy variable for Manhattan and a Manhattan-maximum-gust-speed interaction term:

$$Y_{ij} = \beta_0 + \beta_1 \text{total\_precip}_{ij} + \beta_2 \text{max\_wind}_{ij} + \beta_3 \text{max\_gust}_{ij} + \beta_4 M_{ij} + \beta_5 (\text{max\_gust}_{ij} \times M_{ij}) + u_i + \epsilon_{ij}$$

### 2.5.5 Month and Gust Speed Adjusted Mixed Model for Assessing December Confounding

We built a model exploring significance of month as a factor (so each month is added as a fixed effect) in ridership using random intercepts for origin-destination pair and origin borough, which revealed some months have more significant effects:

$$Y_{ij} = \beta_0 + \sum_{m=2}^{12} \beta_m \times \text{Month}_{m,ij} + u_i + v_k + \epsilon_{ij}$$

According to the earlier plots, December exhibits higher precipitation and maximum gust speeds. Given the effect of gust speed on ridership we saw, it is worth examining whether gust speed is a confounding variable in December's ridership. To do this, we first verified the significance of December on ridership, then used a random intercepts model with the December indicator and maximum gust speed as fixed effects, and origin-destination pair and origin borough as random effects.

$$Y_{ij} = \beta_0 + \beta_1 \times \text{December}_{ij} + \beta_2 \times \text{max\_gust}_{ij} + u_i + v_k + \epsilon_{ij}$$

Adding the maximum gust fixed effect while controlling for origin borough with the random effect yields December losing significance, indicating that maximum gust speed is a confounding variable and may explain the significance of December's effect on ridership. This is confirmed when we fit the same model using the fixed effect for temporal month, e.g. that which was done in Model 2.5.2.

### 2.5.6 Including Principal Components as Fixed Effects

Revisiting the earlier PCA, we introduce the three most important principal components as covariates to account for the effects of weather on ridership. Model 2.2.1 from earlier maintains the random effects for intercept (specific to borough classification) and random effects for slope. The new model includes fixed effects for the interactions between principal components and month. We do not include intercept terms for the PCs to avoid the risk of overfitting the model due to multicollinearity.

$$Y_{ij} = M_i \beta_{0M} + M_i b_{0i,M} + (1-M_i)\beta_{0N} + (1-M_i)b_{0i,N} + M_i t_{ij} \beta_{1M} + (1-M_i) t_{ij} \beta_{1N} + P_{1i} t_{ij} \beta_{1P_1} + P_{2i} t_{ij} \beta_{1P_2} + P_{3i} t_{ij} \beta_{1P_3} + t_{ij} b_{1i} + e_{ij}$$

## 3. CONCLUSIONS

The results from this model tell us that intercepts for Manhattan and the rest of NYC (N) are significant, but neither slope is. This aligns with what we can see in the spaghetti plots of the raw data and the average slopes. Model 2.2.1 produces the output:

| Effect | Value | t_value | p_value |
|---|---|---|---|
| M | 5.15 | 4.61 | 0.00 |
| N | 0.92 | 5.78 | 0.00 |
| M:month | -0.02 | -1.30 | 0.19 |
| N:month | -0.01 | -1.10 | 0.27 |

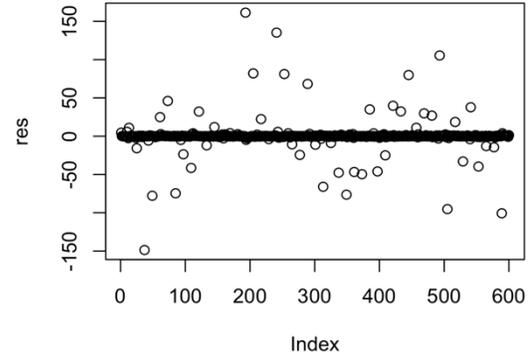

Some of the residuals are very large, but most fall around 0, indicating good fit.

We were originally interested in whether or not there is a linear trend in the data, and whether this varies by borough. The t-tests for the slope parameters were insignificant, telling us that we do not have sufficient evidence for a linear trend for each borough. By including indicators for boroughs in different versions of the model, the output indicates that only the intercept for Manhattan is different from the others. This aligns with what we can see in the plots of the raw data.

Since the analysis with Model 2.2.1 did not provide much insight into the linear trends of the NYC ridership data, we turn toward inclusion of weather components as defined by Model 2.5.6. The output indicates that the intercepts for the different boroughs are significant, as we concluded in the previous part of the analysis. However, we also find that the interaction of PC3 with month is significant. PC3 is a measure of wind behavior, as concluded previously by the loadings. This suggests that wind behavior's relationship with average ridership varies by month.

| Effect | Value | t_value | p_value |
|---|---|---|---|
| M | 5.12 | 4.59 | 0.00 |
| N | 0.89 | 5.61 | 0.00 |
| M:month | -0.01 | -0.92 | 0.36 |
| N:month | -0.00 | -0.49 | 0.63 |
| month:PC1 | 0.00 | 0.07 | 0.94 |
| month:PC2 | 0.00 | 1.82 | 0.07 |
| month:PC3 | 0.01 | 2.16 | 0.03 |

Using the loadings from the PCA defined in 2.4.2, we investigate effects of precipitation, wind speeds, and gust speeds on monthly ridership, starting with Model 2.5.1.

| Effect | Estimate | t_value |
|---|---|---|
| (Intercept) | 2.43 | 5.08 |
| total_precip | 0.00 | 0.35 |
| max_wind | 0.00 | 0.31 |
| max_gust | -0.01 | -2.72 |

This model shows high random effects intercept variability ($\sigma^2 = 10.463$), indicating baseline ridership has high variability between origin-destination pairs. Notice total precipitation and maximum wind speed both are insignificant and have small coefficient estimates, so we remove them from the model and introduce origin borough as a random effect to address bias caused by origin borough. Model 2.5.2 has high random effects intercept variance ($\sigma^2 = 7.303$) and produces

| Effect | Estimate | t_value |
| --- | --- | --- |
| (Intercept) | 2.38 | 2.22 |
| max_gust | -0.01 | -3.13 |
| month | -0.01 | -2.15 |

This shows maximum gust speed and month are both significant, indicating maximum gust speed may have an effect on ridership. Model 2.5.3 attempts to identify where specifically maximum gust speed has an effect on ridership. Maximum gust speed was not significant for any borough except Manhattan, as shown by the model specific to Manhattan-originated trips:

| Effect | Estimate | t_value |
| --- | --- | --- |
| (Intercept) | 6.05 | 4.33 |
| max_gust | -0.02 | -2.52 |

Residual variance in this model was high ($\sigma^2 = 25$). This indicates Manhattan has high sensitivity to gust speeds– when gusts are strong, people may opt out of using the subway in favor of using a taxi or rideshare, or stay home entirely. In addition, people located in surrounding boroughs may be more dependent on transit, or there may be more economic disparity (e.g. commuters in surrounding boroughs may opt for subway usage since travel distance between from other boroughs is generally longer, thus more expensive).

Gust sensitivity may be generally explained by the density of high-rises in Manhattan, which could influence the strength of wind gusts relative to Manhattan– which is congruent with aerodynamic phenomena such as downwash effects, Venturi effect, and vortex formation [6],[7],[8]. Given this reality, we look for interaction between Manhattan and maximum gust speed. Model 2.5.4 expands on this by verifying interaction between Manhattan-originated trips and maximum gust speed, producing the output

| Effect | Estimate | t_value |
| --- | --- | --- |
| (Intercept) | 1.03 | 2.22 |
| total_precip | 0.00 | 0.44 |
| max_gust | -0.00 | -1.24 |
| manhattan_origin | 5.02 | 5.71 |
| max_gust:manhattan_origin | -0.02 | -3.07 |

The significance of the interaction term confirms sensitivity to gusts within the Manhattan region. There were no high correlations ($\rho > 0.5$) between fixed effects. Model 2.5.2 explores the sensitivity to gusts directly.

Month and maximum gust speed effects remain significant and are weakly correlated. Furthermore, AIC comparison of the current model (AIC = 971.784) shows improvement from the baseline random intercepts model (2.5.1) that included fixed effects for total precipitation (AIC = 995.87), maximum wind speed, and maximum gust speed.

Therefore, given the findings from 2.5.2, we conclude Manhattan has sensitivity to wind gust speeds, and wind gust speeds explain ridership variability seen across the year and, more specifically, December. Although this finding is revealing, it is still one small portion of several confounding explanatory variables involved in subway ridership.

Given that NYC heavily relies on public transportation, it makes sense that weather and month would have minimal influence on ridership. Many subway stops are strategically located just beyond a comfortable walking distance, so the choice to ride the subway may remain the same regardless of the weather. It would be interesting to further explore the role of tourism or spatial building volume, but since it peaks during summer and over the winter holidays, the twelve-month setup and linear analysis may not capture this.

# 4. APPENDIX
## 4.1 ETL Pipeline [9]

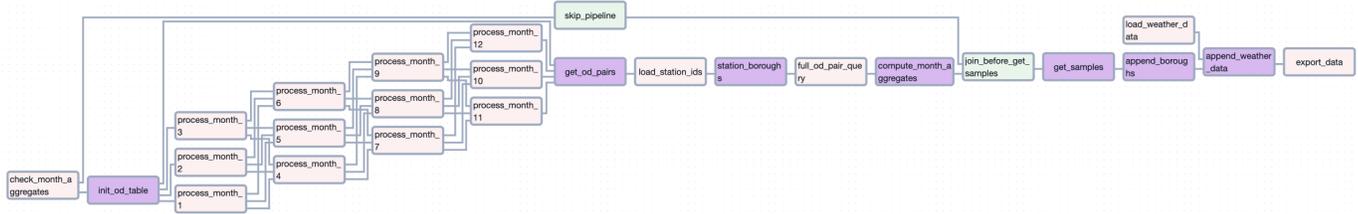

This diagram shows the pipeline represented as a directed acyclic graph (DAG) that was built in Airflow [5]. This DAG builds a complete dataset by pulling, processing, and enriching MTA subway origin-destination (OD) data for 2023. It first checks if a pre-aggregated table already exists; if not, it pulls raw records from the MTA API in 4x3 monthly batches to comply with rate limits, stores them in the PostgreSQL database, and constructs OD pairs, which act as unique subject identifiers. It enriches these OD pairs with station metadata, downloads detailed OD ridership data for each pair, and aggregates monthly averages. From there, it samples 50 OD pairs that have data across all 12 months, appends station names and boroughs, merges in monthly weather data, and classifies each record by season. Finally, it exports the cleaned and enriched dataset as a CSV for downstream analysis, which had the following variables: *origin_id, destination_id, month, avg_ridership, origin_borough, destination_borough, origin_name, destination_name, max_temp, avg_temp, min_temp, max_dew_point, avg_dew_point, min_dew_point, total_precip, max_wind, avg_wind, max_gust, avg_gust, season*. Due to the API rate limitations and indexing of MTA's database, the full pipeline can take as long as 30 minutes to run. To avoid having to re-run the full pipeline during testing, a skip task was added, conditional on whether the pre-aggregated table exists. If the table already exists, we skip all tasks up to get_samples, where we randomly sample OD pairs.

URL https://github.com/hainesdata/subway-ridership-longitudinal-analysis/blob/master/airflow/dags/dag.py.

## 4.2 R Code (1)

```r
library(tidyverse)
library(nlme)
library(ggplot2)
library(patchwork)

data<-read.csv("~/Downloads/ridership_weather.csv")
data$pair_id <- paste(data$origin_id, data$destination_id, sep = "_")

#exploratory plots
ggplot(data = data, aes(x = month, y = avg_ridership, group=interaction(origin_id, destination_id)))+geom_point()+geom_line()

#by destination borough
ggplot(data = data, aes(x = month, y = avg_ridership, group=interaction(origin_id, destination_id), color=destination_borough))+geom_point()+geom_line()

#by origin borough
ggplot(data = data, aes(x = month, y = avg_ridership, group=interaction(origin_id, destination_id), color=origin_borough))+geom_point()+geom_line()+ylim(0,22)+
  scale_x_continuous(breaks = 1:12)+
  labs(x="Month", y="Average Ridership", title="Average Ridership per Trip in 2023 \n (Grouped by Origin Borough)", color="Origin Borough")+
  scale_color_discrete(
    labels = c("Brooklyn", "Bronx", "Manhattan", "Queens"))+
  theme( plot.title=element_text(hjust=0.5))

#plot averages for each origin borough on same plot
avplot<-data%>%
  group_by(origin_borough, month)%>% summarise(mean_ridership = mean(avg_ridership))%>%
  ggplot(aes(x = month, y = mean_ridership, color=origin_borough))+geom_point()+geom_line()+scale_x_continuous(breaks = 1:12)+
  labs(x="Month", y="Average Ridership", title="Average Ridership per Trip \n in 2023 for Queens", color="Origin Borough")+
  theme( plot.title=element_text(hjust=0.5)) + scale_color_discrete(
    labels = c("Brooklyn", "Bronx", "Manhattan", "Queens"))
avplot

#introduce indicator variable for borough
data<-data%>%
  mutate(M=ifelse(origin_borough=="M", 1, 0),
```

```r
         Bk=ifelse(origin_borough=="Bk", 1,0),
         Bx=ifelse(origin_borough=="Bx", 1,0),
         Q=ifelse(origin_borough=="Q", 1, 0))

#model with just month and borough
# mb1_model<-lme(
#   (avg_ridership) ~ -1 + M + Bk + Bx + Q + M:month + Bk:month + Bx:month + Q:month,
#   random = ~  -1 + M + Bk + Bx + Q + month | pair_id,
#   correlation = corCompSymm(form = ~ month | pair_id),
#   data = data,
#   method = "REML")
# summary(mb1_model)
#wont converge

#simplified structure
data$N<-ifelse(data$origin_borough=="M", 0, 1)
#double check that classifications  aren't overlapping
data$M <- ifelse(data$origin_borough == "M", 1, 0)
data$N <- ifelse(data$origin_borough != "M", 1, 0)

mb2_model<-lme(
  (avg_ridership) ~ -1 + M + N + M:month + N:month,
  random = ~  -1 + M + N + month | pair_id,
  correlation = corCompSymm(form = ~ month | pair_id),
  data = data,
  method = "REML")
summary(mb2_model)

#test model 2 with diagonal covariance structure
mb2b_model<-lme(
  (avg_ridership) ~ -1 + M + N + M:month + N:month,
  random = list(pair_id = pdDiag(form = ~ -1 + M + N + month)),
  correlation = corCompSymm(form = ~ month | pair_id),
  data = data,
  method = "REML")
summary(mb2b_model)
AIC(mb2_model,mb2b_model) # general D matrix fits better

# or keep all boroughs just have general random effects
mb3_model<-lme(
  (avg_ridership) ~ -1 + M + Bk + Bx + Q + M:month + Bk:month + Bx:month + Q:month,
  random = ~  month | pair_id,
  correlation = corCompSymm(form = ~ month | pair_id),
  data = data,
  method = "REML")
summary(mb3_model)

# test with diagonal covariance structure too
mb3b_model<-lme(
  (avg_ridership) ~ -1 + M + Bk + Bx + Q + M:month + Bk:month + Bx:month + Q:month,
  random = list(pair_id = pdDiag(form = ~ month)),
  correlation = corCompSymm(form = ~ month | pair_id),
  data = data,
  method = "REML")
summary(mb3b_model) #tells me all the intercepts are signfiicant, does not match with the plot

AIC(mb3_model, mb3b_model) #lower AIC with general covariance structure again, keep general

#obtain residuals for mb2 model
res <- resid(mb2_model, level = 1,  type = "normalized")
res[1:4]
plot(res)

#select weather variables
library(glmnet)
X <- model.matrix(~ max_temp + avg_temp + min_temp + max_dew_point + min_dew_point + avg_dew_point + to-
tal_precip + max_wind + avg_wind+ max_gust + avg_gust, data = data)
X_scaled <- scale(X[, -1])
y <- data$avg_ridership
y <- data$avg_ridership

# PCA to reduce weather covariates
# Check how well month explains variation in weather covariates
weather_vars <- c("max_temp", "avg_temp", "min_temp",
                  "max_dew_point", "avg_dew_point", "min_dew_point",
                  "total_precip", "max_wind", "avg_wind",
                  "max_gust", "avg_gust")

# Create monthly summary (distinct rows per month)
```

```r
monthly_weather <- data[!duplicated(data$month), c("month", weather_vars)]

# Correlation matrix
cor_matrix <- cor(monthly_weather[ , weather_vars])
print(round(cor_matrix, 2))

# PCA to check dominant patterns
pca <- prcomp(monthly_weather[ , weather_vars], scale. = TRUE)
summary(pca)
#PC1 explains about 65% of the variation, PC2 19%, and PC3 only 6%

# Get the first few principal components
monthly_pca_scores <- predict(pca)[ , 1:3]

# Add back to full data on month
monthly_pca_scores_df <- data.frame(month = monthly_weather$month,
                                    PC1 = monthly_pca_scores[, 1],
                                    PC2 = monthly_pca_scores[, 2],
                                    PC3 = monthly_pca_scores[, 3])

# Merge with full dataset
data <- merge(data, monthly_pca_scores_df, by = "month")

# Loadings for PC1, PC2, PC3
round(pca$rotation[, 1:3], 2)

#now include the principal components in the model with interaction
mb2w_model <- lme(
  avg_ridership ~ -1 + M + N + M:month + N:month + PC1:month + PC2:month + PC3:month,
  random = ~ -1 + M + N + month | pair_id,
  correlation = corCompSymm(form = ~ month | pair_id),
  data = data,
  method = "REML")
summary(mb2w_model)
#PC intercepts and slopes not significant
summary(mb2w_model)$varFix

cor_matrix <- cor(data[, c("month", "PC1", "PC2", "PC3")])
print(cor_matrix)
```

## 4.3 R Code (2)

```r
# Load data and add unique ID column
data <- read.csv('data/ridership_weather.csv')
data$pair_id <- paste(data$origin_id, data$destination_id, sep = "_")

library(dplyr)

# Verify no missing records
data %>%
  mutate(pair_id = paste(origin_id, destination_id, sep = "_")) %>%
  group_by(pair_id) %>%
  summarise(n_months = n_distinct(month)) %>%
  summary()

# Check within-subject variation
library(ggplot2)

# Check overall variation visually since too many subjects for individual visual inspection
data %>%
  mutate(pair_id = paste(origin_id, destination_id, sep = "_")) %>%
  group_by(month) %>%
  summarise(
    mean_ridership = mean(avg_ridership),
    sd_ridership = sd(avg_ridership)
  ) %>%
  ggplot(aes(x = month, y = mean_ridership)) +
  geom_line(color = "red", linewidth = 1) +
  geom_ribbon(aes(ymin = mean_ridership - sd_ridership,
                  ymax = mean_ridership + sd_ridership),
              fill = "orange", alpha = 0.2) +
  labs(title = "Mean Profile of Average Ridership",
       y = "Mean Average Ridership", x = "Month")

# analyze max vs avg patterns
library(lme4)

# Model comparing max gust speed and avg gust speed
model_gusts <- lmer(
```

```r
  avg_ridership ~ max_gust + avg_gust + (1 | pair_id),
  data = data
)

# View the model summary
summary(model_gusts)

# PCA to reduce weather covariates, Check how well month explains variation in weather covariates
weather_vars <- c("max_temp", "avg_temp", "min_temp",
                  "max_dew_point", "avg_dew_point", "min_dew_point",
                  "total_precip", "max_wind", "avg_wind",
                  "max_gust", "avg_gust")

# Create monthly summary (distinct rows per month)
monthly_weather <- data[!duplicated(data$month), c("month", weather_vars)]

# Correlation matrix
cor_matrix <- cor(monthly_weather[ , weather_vars])
print(round(cor_matrix, 2))

# PCA to check dominant patterns
pca <- prcomp(monthly_weather[ , weather_vars], scale. = TRUE)
summary(pca)

# Get the first few principal components
monthly_pca_scores <- predict(pca)[ , 1:3]

# Add back to full data on month
monthly_pca_scores_df <- data.frame(month = monthly_weather$month,
                                    PC1 = monthly_pca_scores[, 1],
                                    PC2 = monthly_pca_scores[, 2],
                                    PC3 = monthly_pca_scores[, 3]
)

# Merge with full dataset
data <- merge(data, monthly_pca_scores_df, by = "month")

 # Loadings for PC1, PC2, PC3
round(pca$rotation[, 1:3], 2)

# check difference in borough effect across month
data$month_factor <- factor(data$month, levels = 1:12, labels = c("Jan", "Feb", "Mar", "Apr", "May", "Jun",
                                                                  "Jul", "Aug", "Sep", "Oct", "Nov", "Dec"),
                   ordered = TRUE)

# visualize weather conditions that might help us decide what to explore.
ggplot(data, aes(x = month_factor, y = total_precip)) +
  stat_summary(fun = mean, geom = "line", group = 1) +
  stat_summary(fun = mean, geom = "point") +
  labs(title = "Monthly Average Total Precipitation",
       x = "Month",
       y = "Average Total Precipitation")

library(patchwork)

# Create plots for max and average gust and dew point
p1 <- ggplot(data, aes(x = month_factor, y = max_gust)) +
  stat_summary(fun = mean, geom = "line", group = 1) +
  stat_summary(fun = mean, geom = "point") +
  labs(title = "Monthly Maximum Gust Speed",
       x = "Month",
       y = "Maximum Gust Speed")

p2 <- ggplot(data, aes(x = month_factor, y = avg_gust)) +
  stat_summary(fun = mean, geom = "line", group = 1) +
  stat_summary(fun = mean, geom = "point") +
  labs(title = "Monthly Average Gust Speed",
       x = "Month",
       y = "Average Gust Speed")

p3 <- ggplot(data, aes(x = month_factor, y = max_dew_point)) +
  stat_summary(fun = mean, geom = "line", group = 1) +
  stat_summary(fun = mean, geom = "point") +
  labs(title = "Monthly Maximum Dew Point (Humidity)",
       x = "Month",
       y = "Maximum Dew Point")

p4 <- ggplot(data, aes(x = month_factor, y = avg_dew_point)) +
  stat_summary(fun = mean, geom = "line", group = 1) +
```

```r
  stat_summary(fun = mean, geom = "point") +
  labs(title = "Monthly Average Dew Point (Humidity)",
       x = "Month",
       y = "Average Dew Point")

# arrange in grid
(p1 | p2) / (p3 | p4)

library(lme4)
# check whether precipitation has meaningful effect on ridership, after accounting for random pair effects
model1 <- lmer(avg_ridership ~ total_precip + (1 | pair_id), data = data)
summary(model1)

# add wind, gust, origin, destination covariates
model2 <- lmer(avg_ridership ~ total_precip + max_wind + (1 | pair_id), data = data)
summary(model2)
model3 <- lmer(avg_ridership ~ total_precip + max_wind + max_gust + (1 | pair_id), data = data)
summary(model3)
model4 <- lmer(avg_ridership ~ max_gust + month + origin_borough + destination_borough + (1 | pair_id), data = data)
summary(model4)

# use origin borough as random effect to address bias
model5 <- lmer(avg_ridership ~ max_gust + month + (1 | pair_id) + (1 | origin_borough), data = data)
summary(model5)

# stratify for manhattan origin borough
manhattan_data <- data[data$origin_borough == 'M',]
model6 <- lmer(avg_ridership ~ max_gust + (1 | pair_id), data = manhattan_data)
summary(model6)

# stratify for other boroughs (max gust speed)
queens_data <- data[data$origin_borough == 'Q',]
bronx_data <- data[data$origin_borough == 'Bx',]
brooklyn_data <- data[data$origin_borough == 'Bk',]
model6_q <- lmer(avg_ridership ~ max_gust + (1 | pair_id), data = queens_data)
model6_bx <- lmer(avg_ridership ~ max_gust + (1 | pair_id), data = bronx_data)
model6_bk <- lmer(avg_ridership ~ max_gust + (1 | pair_id), data = brooklyn_data)
summary(model6_q)
summary(model6_bx)
summary(model6_bk)

# Create manhattan dummy variable and add interaction term to mixed model
data$manhattan_origin <- ifelse(data$origin_borough == "M", 1, 0)
model_interaction <- lmer(avg_ridership ~ total_precip + max_gust * manhattan_origin + (1 | pair_id), data = data)
summary(model_interaction)

# In only brooklyn, max humidity (e.g. humidity) has significant negative effect on ridership
# other boroughs do not exhibit max humidity significance
model7_bk <- lmer(avg_ridership ~ max_dew_point + (1 | pair_id), data = brooklyn_data)
summary(model7_bk)

# reveals some months may be more significant than others...
summary(lmer(avg_ridership ~ factor(month) + (1 | pair_id) + (1 | origin_borough), data=data))

# so let's add dummy variable for december, and can visually see december has higher precipitation and max gust
data$december <- ifelse(data$month == 12, 1, 0)

# adding maximum gust to december model while controlling for origin
summary(lmer(avg_ridership ~ december + max_gust + (1 | pair_id) + (1 | origin_borough), data=data))

# this reveals month has significant temporal trend which is partially explained by max gust
model8 <- lmer(avg_ridership ~ month + max_gust + (1 | pair_id) + (1 | origin_borough), data=data)

# add interaction between month and max_gust
model8_b <- lmer(avg_ridership ~ month + max_gust + month*max_gust + (1 | pair_id) + (1 | origin_borough), data=data)

# Test lme_constvar vs model8
AIC(model3, model8)

# model8 w/ pcs
summary(
  lmer(
    avg_ridership ~ month + max_gust + PC2 + PC3 + (1 | pair_id) + (1 | origin_borough) , data=data
  )
)
```

## 5. REFERENCES

[1] Curtis, Z. and Haines, J. Monthly subway ridership with station metadata and weather. URL https://github.com/hainesdata/subway-ridership-longitudinal-analysis/blob/master/data/ridership_weather.csv

[2] Metropolitan Transit Authority. The subway origin-destination ridership dataset. URL https://www.mta.info/article/introducing-subway-origin-destination-ridership-dataset

[3] Weather Underground. Weather History for New York City (Jan 2023 - Dec 2023). URL http://www.wunderground.com/history/monthly/us/ny/new-york-city/KLGA/date/2023-1

[4] Apache Airflow. What is Airflow. *Airflow Documentation*. URL https://airflow.apache.org/docs/apache-airflow/stable/index.html

[5] Apache Airflow. Concepts - DAGs. *Airflow Documentation*. URL https://airflow.apache.org/docs/apache-airflow/1.10.9/concepts.html#:~:text=In%20Airflow%2C%20a%20DAG%20%E2%80%93%20or,and%20their%20dependencies.

[6] Setio, Herlien & Sarli, Prasanti & Sanjaya, Yongky & Priambodo, Doni. (2020). Experimental Study of Wind Flow in a Street Canyon between High-Rise Buildings Using PIV. *Journal of Engineering and Technological Sciences*. 52. 639. 10.5614/j.eng.technol.sci.2020.52.5.3. URL https://www.researchgate.net/publication/346350761_Experimental_Study_of_Wind_Flow_in_a_Street_Canyon_between_High-Rise_Buildings_Using_PIV

[7] Wikimedia Foundation. (2025, April 16). Venturi effect. *Wikipedia*. URL https://en.wikipedia.org/wiki/Venturi_effect

[8] Blocken, B., and S. Roels. "A Numerical Study of Wind Nuisance for a High-Rise Building Group." *Proceedings of the 2nd International Conference on Research in Building Physics, 14-18 September 2003, Leuven, Belgium*, edited by J. Carmeliet, H. Hens, and G. Vermeir, 2003, pp. 981–990. URL https://www.irbnet.de/daten/iconda/CIB2492.pdf

[9] Curtis, Z. and Haines, J. Longitudinal Analysis of 2023 MTA Subway Ridership – Data Workflow Directed Acyclic Graph. URL https://github.com/hainesdata/subway-ridership-longitudinal-analysis/blob/master/airflow/dags/dag.py